\documentclass[prb,superscriptaddress, preprintnumbers,onecolumn,showpacs]{revtex4}
\usepackage{amsmath}
\usepackage{amssymb}
\usepackage{amsthm}
\usepackage{amsfonts}
\usepackage{algorithmic}
\usepackage{enumerate}
\usepackage{latexsym}
\usepackage{graphicx}
\usepackage{bm}
\usepackage{subfigure}
\usepackage{ifthen}
\usepackage{sidecap}
\usepackage{txfonts}
\begin{document}
\title{Route towards Localization for Quantum Anomalous Hall Systems with Chern
Number 2}

\author{Zhi-Gang Song}
\affiliation{SKLSM, Institute of Semiconductors, Chinese Academy of
Sciences, P.O. Box 912, Beijing 100083, China}
\affiliation{Synergetic Innovation Center of Quantum Information and Quantum Physics, University of Science and Technology of China, Hefei, Anhui 230026, China}
\author{Yan-Yang Zhang\footnote{Electronic address: yanyang@semi.ac.cn}}
\affiliation{SKLSM, Institute of Semiconductors, Chinese Academy of
Sciences, P.O. Box 912, Beijing 100083, China}
\affiliation{Synergetic Innovation Center of Quantum Information and Quantum Physics, University of Science and Technology of China, Hefei, Anhui 230026, China}
\author{Jun-Tao Song}
\affiliation{Department of Physics and Hebei Advanced Thin Film Laboratory, Hebei Normal University, Hebei 050024,, China}
\author{Shu-Shen Li}
\affiliation{SKLSM, Institute of Semiconductors, Chinese Academy of
Sciences, P.O. Box 912, Beijing 100083, China}
\affiliation{Synergetic Innovation Center of Quantum Information and Quantum Physics, University of Science and Technology of China, Hefei, Anhui 230026, China}




\begin{abstract}
The quantum anomalous Hall system with Chern number 2 can be destroyed by sufficiently strong disorder.
During its process towards localization, it was found that the electronic states will be directly localized to an Anderson insulator (with Chern number 0), without an intermediate Hall plateau with Chern number 1. Here we investigate the topological origin of this phenomenon, by calculating the band structures and Chern numbers for disordered supercells. We find that on the route towards localization, there exists a hidden state with Chern number 1, but it is too short and too fluctuating to be practically observable. This intermediate state cannot be stabilized even after some ``smart design'' of the model and this should be a universal phenomena for insulators with high Chern numbers. By performing numerical scaling of conductances, we also plot the renormalization group flows for this transition, with Chern number 1 state as an unstable fixed point. This is distinct from known results, and can be tested by experiments and further theoretical analysis.
\end{abstract}

\maketitle
%
%
\thispagestyle{empty}

\section*{Introduction}

As well as integer quantum Hall effect (QHE)
under a magnetic field\cite{Klitzing},
the quantum anomalous Hall effect (QAHE) without an external
magnetic field\cite{Haldane} is characterized by nonzero Chern number,
a topological invariant associated with occupied bands\cite{TKKN,TwistedBC1}.
This nonzero Chern number $C$ gives rise to edge states in the bulk gap,
which carry the remarkably perfect quantization of the Hall conductance $\sigma_{xy}=C$ (in units of $\frac{e^2}{h}$ hereafter).
Inspired by the investigations of topological insulators\cite{Hasan,XLQiRMP,SQShen},
QAHEs in concrete materials have been theoretically proposed \cite{QAH_CXLiu,QAH_HJZhang,QAH_QZWang,QAH_RYu,QAH_SCWu,QAH_ZHQiao} and
been experimentally observed \cite{QAH_Exp1,QAH_Exp2,QAH_Exp3,QAH_Exp4,QAH_Exp5}.

Although robust against weak disorder,
topological phases will be localized into Anderson insulators ultimately, by
sufficiently strong disorder\cite{QHE_XCX,QHE_ZYW,QHE_Disorder,TI_Disorder}.
Microscopically speaking, this can be viewed as a disorder induced topological phase transition,
i.e., change of the topological invariant ($1\rightarrow 0$) due to a band touching, i.e.,
gap closing and re-opening\cite{QHE_Disorder,TI_Disorder}.

Topological phases with different Chern numbers are topologically distinct from each other.
Each of them corresponds to a \emph{stable} fixed point in the sense of renormalization group (RG)\cite{QHE_RG}, which offers another explanation for their robustness.
Numerical calculations for a QHE with a high Chern number $C>1$ show that,
with the increasing of disorder strength,
the Hall conductance $\sigma_{xy}=C$ vanishes to 0 persistently without showing any intermediate Hall plateaus with Chern numbers $C-1,C-2,\cdots,1$\cite{QHE_XCX,QHE_ZYW,QHE_Disorder}, which are also associated with stable fixed points and therefore should be robust.
The quantum Hall state with $C>1$
is ascribed to contributions from individual occupied Landau levels with $C=1$.
Disorder makes the Landau levels broaden.
After adjacent Landau levels are broadened enough to touch each other,
their Chern numbers change from $\pm1$ to $0$, which results in trivial localized bands.
The most essential observation is that,
this disorder induced topological transitions happen to \emph{all} Landau levels almost at
similar disorder strength, which means a nearly simultaneous breakdown of all Landau levels\cite{QHE_XCX,QHE_Disorder}.
Therefore, after disorder ensemble average, any intermediate topological state with only a fraction of the Landau levels surviving cannot stably manifests itself.

Lattice models for QAHE with a single valence band carrying a high Chern number $C>1$ were proposed recently\cite{HighChern1,DSticlet2012,HighChern2}.
The localization process with increasing disorder strength for a two-band model with $C=2$ was investigated by transport calculations \cite{JTSong2015}.
It was found that its Hall conductance still decays from 2 persistently to 0 without a plateau 1.
Now since one single band carries $C=2$, the above picture of simultaneous breakdowns for all
Landau levels (each carrying $C=1$) for QHE\cite{QHE_Disorder} does not apply.
Without special symmetry protection, the band touching induced by a single parameter (disorder strength here)
is linearly shaped and only changes the Chern number by 1\cite{QHE_Disorder,Hatsugai2000,Murakami2007}.
Therefore, the topological origin of above phenomenon seems to be confusing. Does this correspond to a topological phase transition from two band-touching points happening simultaneously, or from a single band-touching point changing the Chern number by 2?

In this manuscript, based on the language of disordered supercells\cite{TwistedBC1,TwistedBC2,QHE_Disorder,TI_Disorder,TAI_YYZ}, we investigate the topological origin of the localization process for QAHE with Chern number 2 $\rightarrow$ 0. We find that,
with the increasing of disorder strength $W$,
this transition corresponds to two successive band-touchings which happen at different disorder strengths $W_1$ and $W_2$,
and at different positions $\bm{k}_1$ and $\bm{k}_2$ in the Brillouin zone (BZ).
The intermediate $C=1$ window $(W_1,W_2)$ is small and highly configuration dependent,
which makes it hardly observable after disorder averaging and size scaling.
This vulnerability of intermediate $C=1$ state cannot be improved even if we generalize
the original model to make the gap at one valley much smaller than at other valleys
(thus with the expectation that the band-touching at the smallest gap should be much earlier). Finally, based on finite-size scaling of mesoscopic conductances,
we summarize the results into a diagram of RG flows, which can be tested by experiments\cite{RG_Exp2,RG_Exp1} or field theoretical analysis\cite{QHE_RG}.

\section*{Results and Discussions}

We use a generalized version of the two-band model recently proposed\cite{DSticlet2012,JTSong2015}.
This is a spinless model defined on a square lattice with two orbitals at each site.
Setting the lattice constant $a=1$, the Hamiltonian in $k$ space reads\cite{JTSong2015}
\begin{align}
H_0(\bm{k})&=d_{1}\sigma _{x}+d_{2}\sigma
_{y}+d_{3}\sigma _{z}  \label{eqH0}  \\
d_{1}(\bm{k})&=2t_{1}\cos k_{x},  \qquad d_{2}(\bm{k})=2t_{1}\cos k_{y},  \notag \\
d_{3}(\bm{k})&= m+2t_{3}\sin
k_{x}+2t_{3}\sin k_{y} \notag + 2t_{2}\cos(k_{x}+k_{y}),  \notag
\end{align}%
where $\sigma_i$ are Pauli matrices in the orbital space.
After realizing the Hamiltonian on the square lattice, one can see that $m$ is the staggered potential on two orbitals,
and $t_i$ ($i\neq 2$) are the nearest hoppings between different sites.
The next nearest hopping $t_2$ just exists along \emph{one} direction, e.g. along two-dimensional crystallographic orientation $[11]$, which is essential for
realizing Chern number 2\cite{DSticlet2012,JTSong2015}.
Hereafter, $t_1=1$ will be used as the energy unit.
In the BZ, model (\ref{eqH0}) has four Dirac points (or valleys) at $D_{1,2,3,4}=\left(  \pm\frac{\pi
}{2},\pm\frac{\pi}{2}\right)$ and associated masses (or gaps) $\Delta_i$ are
$2((m-2t_2)\pm 4t_3)$ and $2(m+2t_2)$, respectively.

The Hall conductance is related to the totol Chern number of all occupied bands as $\sigma_{xy}=\frac{e^2}{h}\sum_{n}c_n$. The Chern number associated with the $n$-th band is defined
in two-dimensional $k$-space:
\begin{align}
c_{n}&=\frac{1}{2\pi i}\int\nolimits_{BZ}
d^{2}\bm{k}F_z(\bm{k}), \label{4a} \\
\bm{F}&=\nabla_k\times\mathbf{A}, \qquad
\mathbf{A}=
\left\langle n(\bm{k})|\nabla_k|n(\bm{k})\right\rangle, \notag
\end{align}
where $\bm{F}$ is the Berry curvature, and $\left|n(\bm{k})\right\rangle$ is the normalized wave function of the $n$-th
Bloch band such that $H(k)\left|n(\bm{k})\right\rangle=E_n(k)\left|n(\bm{k})\right\rangle$.
For our model with Hamiltonian (\ref{eqH0}), the Chern number associated with the valance band is
straightforwardly determined by the model parameters as\cite{DSticlet2012}
\begin{align}
C  &  =\frac{1}{2}[\mathrm{sgn}(m-2t_{2}+4t_{3})+\mathrm{sgn}(m-2t_{2}-4t_{3}%
)\label{7}  -\mathrm{sgn}(m+2t_{2})-\mathrm{sgn}(m+2t_{2})].\nonumber
\end{align}

\begin{figure}[tp]
\begin{center}
\includegraphics*[width=0.6\textwidth,bb=50 0 1150 504]{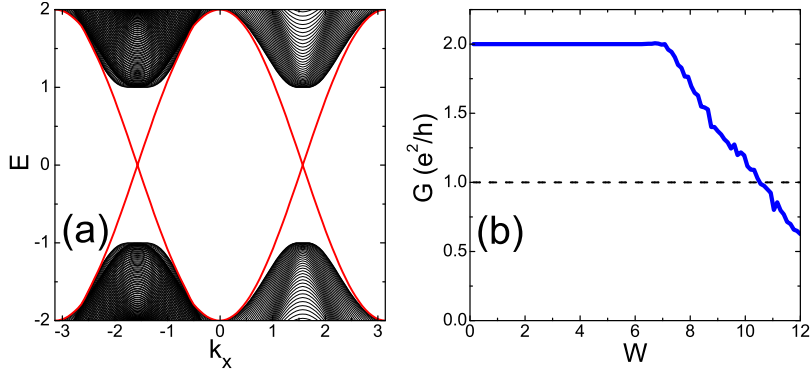}%
\caption{(a) The band structure of the ribbon in clean limit with width $L_y=100$.
(b) Two-terminal conductance as a function of disorder strength, which is averaged over 500 disorder configurations with sample size $L_x=L_y=100$. The model parameters as follows:
$t_{1}=1.0$, $t_{2}=1.0$, $t_{3}=0.5$, $m=-1.0$, and $E_{F}=0.1$.}%
\label{Fig1}%
\end{center}
\end{figure}

The effect of disorder is included in the real
space representation by adding a random potential term $U_i$ to each site $i$,
with $U_i$ uniformly distributed in the interval $(-W/2,W/2)$,
where $W$ is the disorder strength. The Hall conductance of a disordered sample can be simulated by using the standard method of non-equilibrium Green's functions\cite{Datta}. On the other hand, disorder breaks the translation symmetry of the original lattice
and makes the original $k$ space badly defined for Chern number in equation (\ref{4a}).
Nevertheless, for a disordered sample with size $L_x \times L_y$,
if twisted boundary conditions $\exp (i k_x L_x)$ and $\exp (i k_y L_y)$
are introduced in both directions, the definitions of $\bm{k}=(k_x,k_y)$ and Chern number
can be restored\cite{TwistedBC1,TwistedBC2}. This is not surprising since this is equivalent to constructing
a superlattice with this $L_x \times L_y$ sample as the supercell\cite{TI_Disorder,TAI_YYZ}.
Physically reliable results for ``really'' disordered systems will be recovered
after disorder ensemble averaging and size scaling $L\rightarrow \infty$.
See Methods for details of these calculations.

We first reproduce the main results of previous transport calculations\cite{JTSong2015}.
Fig. \ref{Fig1} (a) is the band structure for a quasi-one dimensional ribbon,
where two pairs of edge states from $C=2$ can be seen.
Although there are essentially 4 massive Dirac points for the two-dimensional band structure,
only 2 of them are distinguishable here due to the cutting orientation of producing the quasi-one dimenional ribbon.
In Fig. \ref{Fig1} (b),
we display the the averaged two-terminal conductance $G$ ($=\sigma_{xy}$ when $C\neq 0$) as a function of disorder strength $W$.
The conductance of the $C=2$ state is perfectly quantized regardless
of finite disorder, until $W\sim 6$.
Afterwards, as found previously, the conductance decays from $2$ towards $0$
persistently, without any plateau at the $C=1$ state (the dashed line)\cite{JTSong2015}.

In order to reveal the topological origin of the above transport results,
we resort to the calculations of band structures and Chern numbers
by using the disordered supercell method introduced in Methods.
Random potential
$U_i=W\cdot u_i$ are assigned to each site $i$ of a supercell with $L_x\times L_y$ sites, where $u_i$ are
random numbers uniformly distributed within the unit interval $(-0.5,0.5)$, and $W$ is the disorder strength.
Hereafter, $\{U_i\}=W\cdot u_i$ from a fixed configuration of $\{u_i\}$
but only differed by the factor $W$ will still be called \emph{one} disorder configuration (or one sample).
Applying twisted boundary conditions to this disordered supercell,
we can investigate the adiabatic changes of band structures and Chern numbers
with the increasing of the single parameter $W$, for a fixed disorder configuration $\{u_i\}$.

\begin{figure*}[tp]
\begin{center}
\includegraphics*[width=0.32\textwidth,bb=35 0 1120 1350]{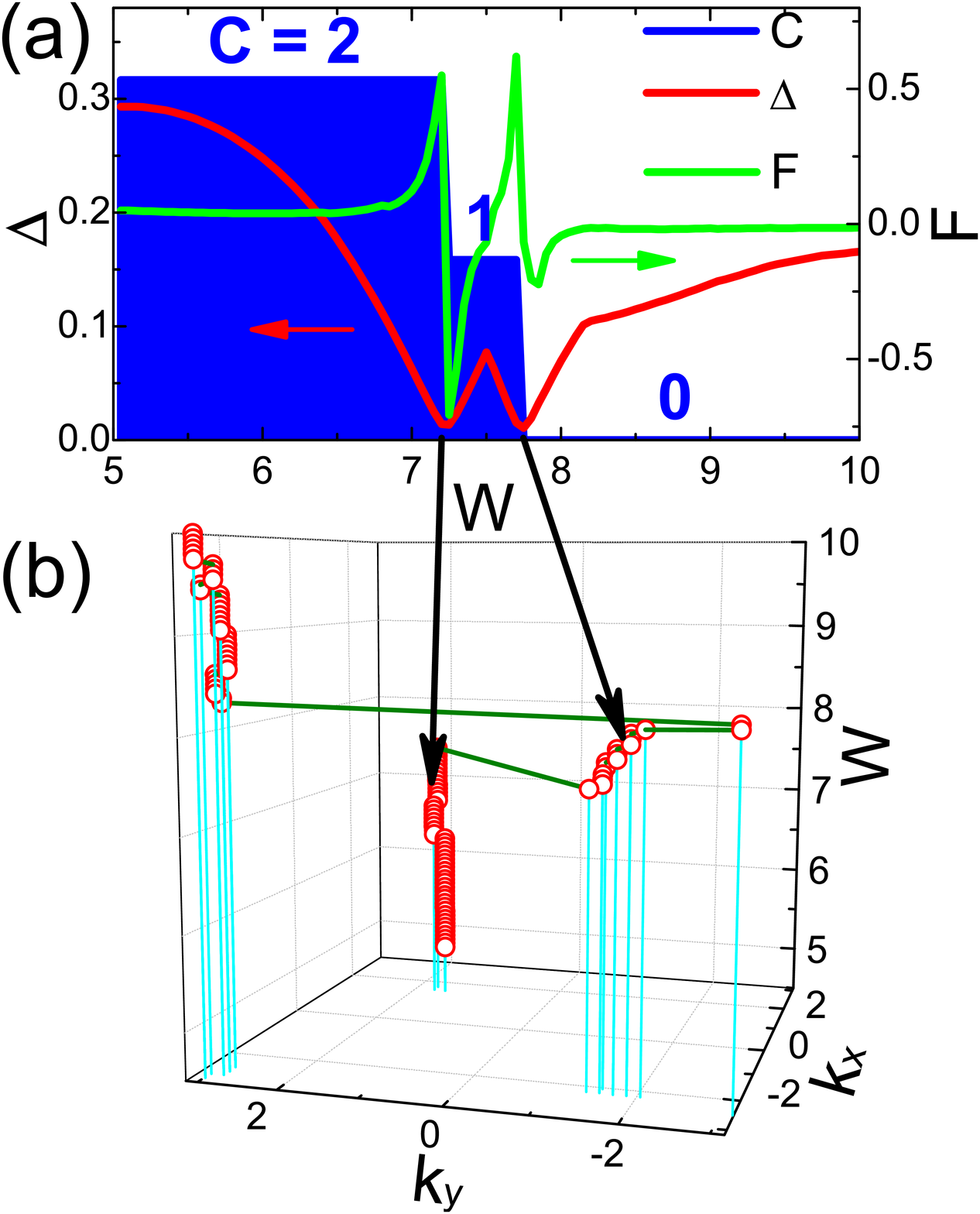}
\includegraphics*[width=0.32\textwidth,bb=35 0 1120 1350]{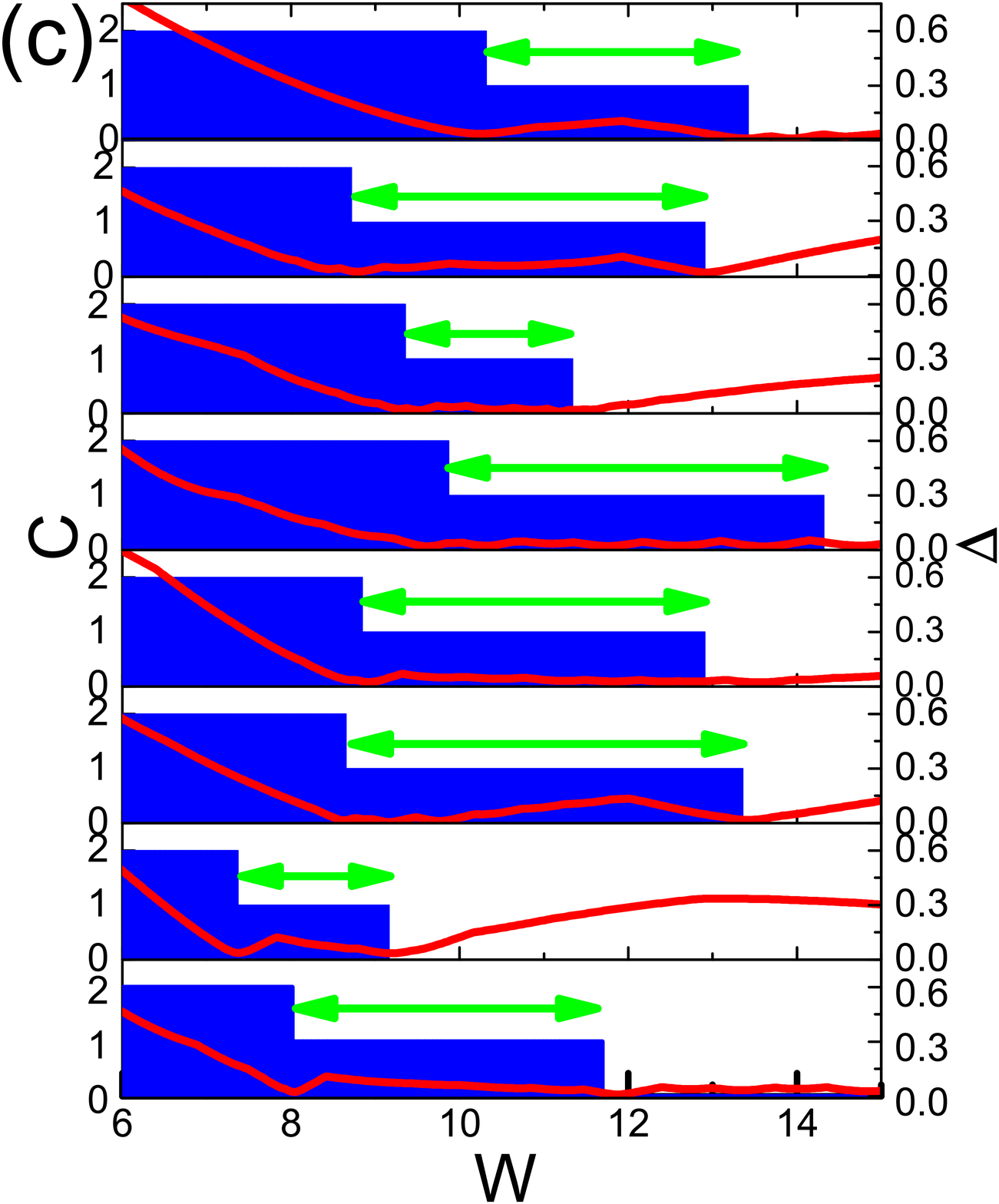}
\caption{(a) Three quantities as functions of disorder strength $W$ for a definite disorder configuration: the band gap $\Delta$ (red line), the Berry curvature $F$ (green line) and the Chern number $C$ of the valence band (blue bar). (b) The location of minimal gap point $k_{\mathrm{min}}$ (red circles) as a function of disorder strength $W$ for the same disorder configuration. The cyan lines indicate their projections on the $k_x-k_y$ plane. (c) The band gap $\Delta$ (red line) and the Chern number $C$ below the gap (blue bar) as functions of disorder strength for 8 different disorder configurations. The intermediate windows $(W_1,W_2)$ with $C=1$ are marked by green arrows.}
\label{Fig2}%
\end{center}
\end{figure*}

Let's first focus on the microscopic evolution of one typical disorder configuration.
To find possible band touching, we define the band gap $\Delta$ as the minimal difference between conduction and valence bands $E_{c(v)}(\bm{k})$ (with the same $\bm{k}$) in the BZ,
\begin{equation}
\Delta\equiv \min_{\bm{k}\in \mathrm{BZ}}\{E_c(\bm{k})-E_v(\bm{k})\},\nonumber
\end{equation}
with the associated point $k_{\mathrm{min}}$ as the minimal gap point. A band touching corresponds to $\Delta=0$. At the clean limit, $k_{\mathrm{min}}$ is just one of the Dirac points $D_i$.
In Fig. \ref{Fig2} (a), we plot the band gap $\Delta$ (red line), the Berry curvature $F$ at the minimal gap point $k_{\mathrm{min}}$ (green line) and the Chern number $C$ of the valance band (blue bar) as functions of disorder strength $W$, around the transition region.
It is interesting to notice that for a definite configuration,
there are \emph{two} subsequent topological transitions at $W_1\sim 7.2$ and $W_2\sim 7.7$, with band touchings ($\Delta=0$), and with Chern numbers
$2\rightarrow 1$, then $1\rightarrow 0$, respectively.
In Fig. \ref{Fig2} (b), we plot the track of the minimal gap point $k_{\mathrm{min}}$ in the BZ as a function of disorder strength $W$. Approaching the first topological transition $W_1\sim 7.2$ from below, $k_{\mathrm{min}}$ begins to digress from Dirac point $D_1=(\frac{\pi}{2},\frac{\pi}{2})$. Soon after this, $k_{\mathrm{min}}$ appears around another Dirac point $D_3=(-\frac{\pi}{2},-\frac{\pi}{2})$, and the second topological transition $W_1\sim 7.7$ happens around here. Such large transfer of the minimal gap point is not surprising since the topological phase collapses at strong disorder, which has remarkably changed the original valley shapes in the clean limit.

After confirming the existence of an intermediate $C=1$ state for a definite disorder configuration,
we have to answer why it cannot be observed in the transport calculations in Fig. \ref{Fig1} (b).
In Fig. \ref{Fig2} (c), we show the gaps $\Delta$ (red curve) and Chern numbers $C$ (blue bar) as functions of $W$ for 8 different disorder configurations, each with an intermediate $C=1$ state between $W_1$ and $W_2$ (the green arrows). The first observation is that these $C=1$ windows are rather narrow, i.e., $W_2-W_1\ll W_1$ (Note the leftmost of the $W$-axis is $6$ instead of $0$.).
Another important observation is that such windows
$(W_1,W_2)$ with $C=1$ are quite configuration dependent, with their locations and lengths fluctuating remarkably
from configuration to configuration. There is not a stable
common window with Chern number $C=1$ for \emph{all} different configurations. Furthermore, we have checked that,
this absence of a stable common window for $C=1$ does not get better at all,
with the increasing of sample size $L$. In short, although there exists an
intermediate window of $C=1$ for each definite configuration,
this window is too vulnerable to survive after disorder averaging and size scaling,
and thus is physically unobservable.
This is the first important finding
in this manuscript.

\begin{figure}[tp]
\includegraphics*[width=0.6\textwidth,bb=50 0 1150 504]{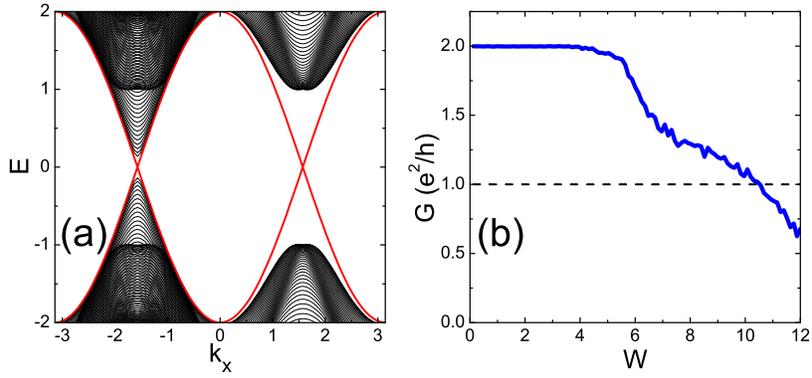}%
\caption{ The same with Fig. \ref{Fig1} but with model parameters as $t_{1}=1.0$, $t_{2}=1.0$, $t_{3}=-0.72$, $m=-1.0$, and $E_{F}=0.1$. (a) The band structure of the ribbon in clean limit.
(b) The disorder averaged two-terminal conductance as a function of disorder strength.}%
\label{Fig3}%
\end{figure}

The two subsequent band touchings at $W_1$ and $W_2$ are too close to
make up a stable state with $C=1$ within $(W_1,W_2)$.
One may blame this to the almost identical gap sizes at two Dirac points in the clean limit
of the model (See Fig. \ref{Fig1} (a)).
This tempts us to modify the model to make the
gap at one Dirac point, say, $\Delta_1$ much smaller than those at other Dirac points.
This can be realized for model (\ref{eqH0}) by setting parameters as depicted in Fig. \ref{Fig3}.
Then, under perturbation from rather weaker disorder\cite{TAI_YYZ},
we expect the first band touching $W_1$ to occur
around this Dirac point $\Delta_1$ with the narrowest gap, much earlier than $W_2$ occurring
around another Dirac point. Thus a wider window $(W_1,W_2)$ could be expected and this
could be beneficial for stabilizing the $C=1$ even after
disorder averaging.

In Fig. \ref{Fig3} (a), the band structure for a ribbon
is plotted. The gap at $D_1$ is 0.24, only 4\% of that at $D_2$.
The two-terminal conductance as a function of increasing disorder is plotted in
Fig.\ref{Fig3} (b). Compared to Fig. \ref{Fig1} (b), the $C=2$ plateau starts to collapse earlier at $W\sim 4$, which can be attributed to the small gap at the clean limit.
However, a stable plateau of $C=1$ is still absent.

\begin{figure}[bp]
\includegraphics[width=0.4\textwidth]{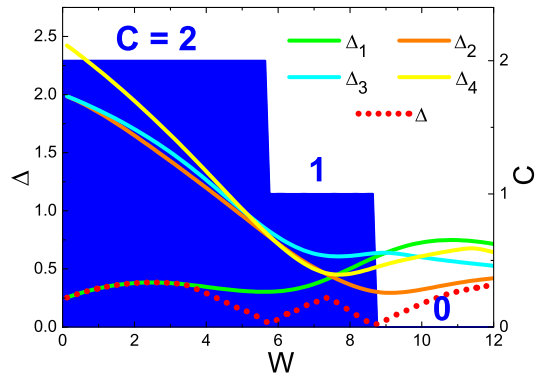}%
\caption{With model parameters same as in Fig. \ref{Fig3}, the developments of quantities for a disorder configuration as functions of disorder strength $W$: the local gaps at 4 Dirac points $\Delta_i$ (solid lines), the minimal band gap $\Delta$ (red dotted line) and the Chern number $C$ (blue bar). Note that after $W>3.6$, the minimal band gap will not be located at any of the Dirac points any more. }
\label{Fig4}%
\end{figure}

We repeat the calculations of disordered supercells to reveal the
underlying topological nature. In order to see the evolutions of different massive Dirac points, local gaps at 4 Dirac points $D_i$ are simply defined as
\begin{equation}
\Delta_i\equiv E_c(D_i)-E_v(D_i).\nonumber
\end{equation}
Without disorder, $\Delta_1$ is just the band gap $\Delta$,
which is defined at the narrowest point in the gap.
In Fig. \ref{Fig4}, we plot the developments of $\Delta_i$ (solid lines),
as well as $\Delta$ (red dotted line) when increasing disorder strength $W$ for a typical disorder configuration.
The Chern number (blue bar) is also plotted as a reference.
In clean limit $W=0$, as we set, $\Delta_1$ (green line) [which is equal to $\Delta$ (red dotted line)] is much smaller than other 3 local gaps.
With the increasing of $W$, those 3 large gaps decrease.
Contrary to what we are expecting, the smallest gap $\Delta_1$ is increasing
to prevent itself from an early band touching at weak disorder.
After $W\sim 3.6$, $\Delta$ does not follow $\Delta_1$ (nor any other $\Delta_i$) any more,
which means the minimal gap point has drifted away from the Dirac points.
This, as discussed above, is a consequence of remarkable destroy of the original
shape of the band structure from strong disorder.
In other words,
the elaborately designed band structure with extremely small local gap
at one Dirac point $\Delta_1$ has been smeared out completely by disorder, before the first band touching occurs.
Then, as happened in Fig. \ref{Fig2}, there will be two subsequent and close band touchings,
which fluctuate so seriously that the $C=1$ state between them is unobservable after
disorder averaging and size scaling. In one word, in the BZ, states around all 4 Dirac points seem to ``feel'' each other and tend to evolve cooperatively under disorder. This reflects
the fact that the Chern number $C$ and its transition are global properties of the band\cite{Hasan,XLQiRMP,TwistedBC1},
especially in the presence of disorder, i.e., without constraints from most spacial symmetries.
We conjecture that
for Chern insulators with high Chern numbers $C>1$,
either QHEs with multiple Landau levels or QAHEs with a single valence band,
this route towards localization under disorder with Chern number $C\rightarrow 0$ is universal,
independent of the concrete details of the materials in the clean limit.

\begin{figure}[tp]
\begin{center}
\includegraphics[width=0.6\textwidth]{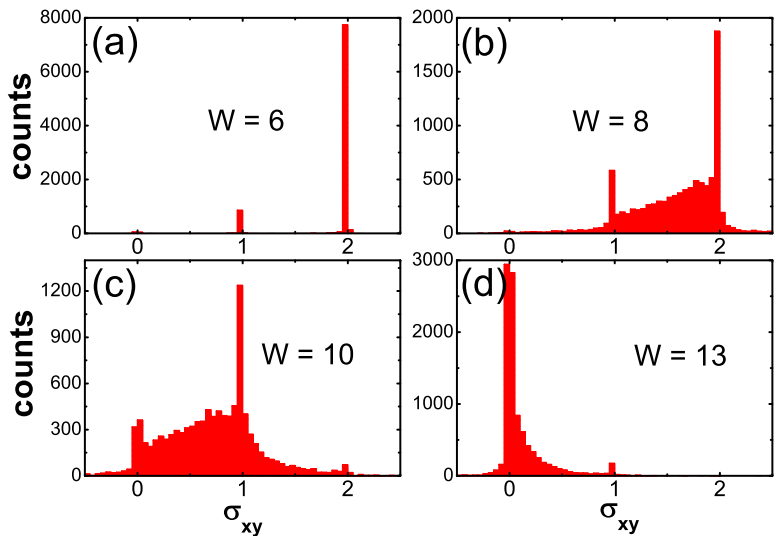}
\caption{Distributions of $(\sigma_{xy}$ for 10000 disorder configurations with supercell size $20\times 20$, for different disorder strengths $W$. The Fermi energy is fixed at $E_F=-0.053$, near the gap centre of the clean limit. Other model parameters are same with Figs. \ref{Fig1}. }%
\label{FigStatistics}%
\end{center}
\end{figure}

It is interesting to revisit this route towards localization, by observing the evolution of the distribution of the Hall conductance. In Fig. \ref{FigStatistics}, we present the statistical histogram of $\sigma_{xy}$ at a definite Fermi energy and sample size, for different disorder strength $W$. We will see that they also reflects the physical pictures described above. For $W=6$ when $\sigma_{xy}$ starts to decrease from 2, Fig. \ref{FigStatistics} (a) shows that a small portion (around 10\%) of samples have experienced a topological transition $2\rightarrow 1$. For a larger $W$ as illustrated in Fig. \ref{FigStatistics} (b), besides two peaks at $\sigma_{xy} = 1, 2$, a considerable amount of samples with non-quantized $\sigma_{xy}$ appear. They  originate from partly filled subbands randomly distributed around the Fermi energy, which are induced by strong disorder. Such a broad distribution gives rise to a non-quantized $\sigma_{xy}$ with a large statistical variation.
When $W=10$ [Fig. \ref{FigStatistics} (c)], most $C=2$ states have collapsed. However, there is a prominent and sharp peak at $C=1$. This means that $E_F=-0.053$ here is within the (small but finite) gaps of these samples with $C=1$ below them; Or in other words, $W=10$ here is within the $C=1$ windows $(W_1,W_2)$ [e.g., those marked by green arrows in Fig.\ref{Fig2} (c)] of these samples.
Nevertheless, as stated earlier, there is no common window with $C=1$ for all (or most) samples. For the rest samples (more than 80\%), their $\sigma_{xy}$ are still widely distributed, making a stable and observable $\sigma_{xy}=1$ state impossible. Fig.\ref{FigStatistics} (a) to (c) reaffirm the ``hidden'' nature of the intermediate $C=1$ state: it is existent but overwhelmed by strong statistical fluctuations, and thus hardly observable. This is contrary to the case of topological Anderson insulator\cite{TAI}. In that case, although also with dense disorder induced in-gap states, the topological invariant still has a probability of almost $100\%$ to be 1, in a finite range of energy and disorder strength, giving rise to an observable quantized conductance plateau with tiny fluctuations\cite{TAI_YYZ}. At the end of this transition, as shown in Fig.\ref{FigStatistics} (d), most samples have crossed the second topological transition $W_2$ towards trivial localization with $C=0$.

\begin{figure}[tp]
\begin{center}
\includegraphics[width=0.7\textwidth]{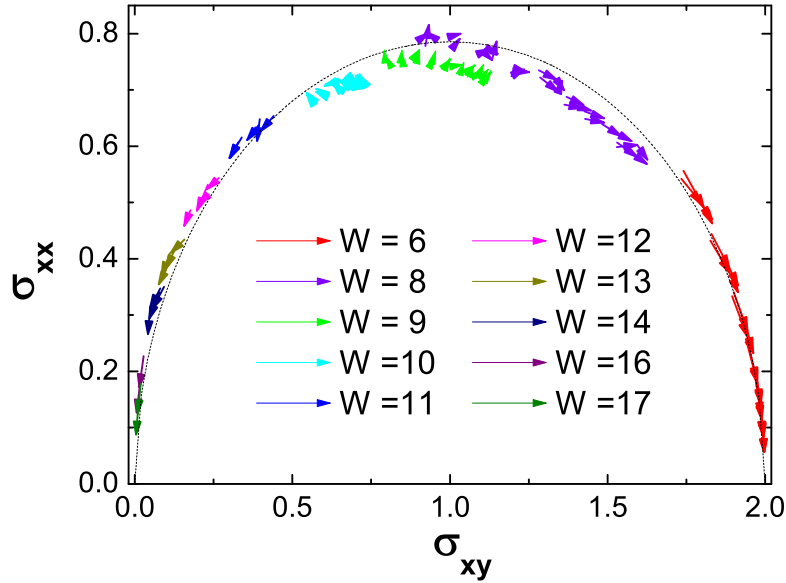}
\caption{The renormalization group flows extracted from finite size scaling for square samples with $L_x=10\rightarrow15\rightarrow20$, with Fermi energies $E_F\in(-1,1)$ and disorder strengths as shown. Other model parameters are same with Fig. \ref{Fig1} and each point is an average over 10000 disorder configurations (50000 configurations for $W=9,10$). The dashed line is the semi-ellipse with the semi-major (semi-minor) axis $a=1$ ($b=\sigma_1$), where $\sigma_1\equiv \frac{\pi}{4}$. The $C=1$ state corresponds to point $(1,\sigma_1)$. }%
\label{FigRGFlow}%
\end{center}
\end{figure}

Finally let's put this localization process into an RG flow diagram by numerical scaling. RG analysis can provide vivid and deep understanding of phase transitions. Besides, recent experimental progresses have made RG flows for QAHEs directly observable\cite{RG_Exp2,RG_Exp1}. The sample size driven $(\sigma_{xy},\sigma_{xx})$ flows are plotted in Fig. \ref{FigRGFlow}, corresponding to the localization process in Fig. \ref{Fig1}. Similar to the RG diagram in the transition with $C=1\rightarrow 0$\cite{RG_Exp2,NumScaling,Prodan2013}, most flow arrows reside on a semi-ellipse.
On two ends of the semi-ellipse, scaling flows are attracted to two fixed points $(\sigma_{xy},\sigma_{xx})=(C,0)$, corresponding to the $C=2$ ($C=0$) state before (after) the disorder induced transition, respectively. On the other hand, the intermediate state $C=1$ is located on the uppermost of this semi-ellipse, with the maximum longitudinal conductance $\sigma_1=0.79\sim \pi/4$. Around this point, some scaling flows are running away, and some are randomly oriented. The latter behavior can be attributed to the statistical precision limited by the number of disorder configurations we can achieve, but also reflects the unstable nature of this point. Thus it's reasonable to conclude that this point $(1,\sigma_1)$ is a saddle type \emph{unstable} fixed point, like the role that $(\frac{1}{2},\sigma^*)$ plays in the transition with Chern number $1\rightarrow0$\cite{QHE_RG,NumScaling,Prodan2013}. In brief, we show that the state $C=1$ does not always correspond to a stable fixed point in the sense of RG. Furthermore, the associated maximum longitudinal conductance $\sigma_1=0.79\sim \pi/4$, which is twice as that found in the localization process with Chern number $C=1\rightarrow 0$ state\cite{NumScaling}.
These features make Fig. \ref{FigRGFlow} distinct from known RG knowledge for Chern insulators so far. This can be tested by further theoretical analysis and experiments.

In summary, by investigating the band structures and Chern numbers of disordered supercells, we illustrate the route towards localization for QAHE with $C=2$. With the increase of disorder strength, there are two successive topological transitions for each disorder configuration, but the intermediate $C=1$ is remarkably short and fluctuating, therefore cannot survive after disorder averaging and size scaling. This picture is also supported by statistical distributions and RG flows of the conductances.

\section*{Methods}

At zero temperature, the two-terminal conductance of a finite sample
can be expressed by Green's functions as\cite{Datta}
\begin{equation}
G=\frac{e^{2}}{h}\mathrm{Tr}\left[  \Gamma_{S}G^{r}\Gamma
_{D}G^{a}\right]  , \label{3}%
\end{equation}
where
$G^{r/a}$ is the retarded/advanced Green's function, and $\Gamma_{S(D)}%
=i(\Sigma_{S(D)}^{r}-\Sigma_{S(D)}^{a})$ with $\Sigma_{S(D)}%
^{r/a}$ being retarded/advanced self energies due to the source (drain) lead, respectively.
For a quantum (anomalous) Hall state ($C\neq 0$), this two-terminal conductance is just
the Hall conductance, $G=\sigma_{xy}$\cite{Datta}.
The transport results in Figs. \ref{Fig1} (b) and \ref{Fig3} (b) were calculated by this method.
However, for topologically trivial state, e.g., a state close to localization, this simple two-terminal $G$ may not rigorously equal to $\sigma_{xy}$ since it will contain bulk transports.
Another method of calculating $\sigma_{xy}$ will be introduced below.

Numerical methods are necessary for treating disordered supercells with large size $L_x\times L_y$. However, the numerical calculation of topological quantities is a tricky task: Special care must be taken to ensure the gauge invariance
in the numerical processes. We adopt a method of discretizing equation (\ref{4a}) as\cite{Chern_Num}
\begin{equation}
c_{n}=\frac{1}{2\pi i}\sum\limits_{l}\mathcal{F}(k_{l}). \label{4}%
\end{equation}
Here $l$ runs over all lattices over the discretized BZ.
On this discretized BZ, the Berry curvature $\mathcal{F}(k_{l})$ is a lattice field defined on each small plaquette as $\mathcal{F}%
(k_{l})=\ln \mathcal{U}_{1}(k_{l})\mathcal{U}_{2}(k_{l}+\widehat{1})\mathcal{U}_{1}(k_{l}+\widehat{2}%
)^{-1}\mathcal{U}_{2}(k_{l})^{-1}$, where $\mathcal{U}_{\mu}(k_{l})\equiv\frac{\left\langle
n(k_{l})|n(k_{l}+\widehat{\mu})\right\rangle }{\left\vert \left\langle
n(k_{l})|n(k_{l}+\widehat{\mu})\right\rangle \right\vert }$, and $\widehat{1}$ and
$\widehat{2}$ are unit vectors of of the discretized BZ\cite{Chern_Num}.

For the purpose of plotting RG flows, both the longitudinal conductance $\sigma_{xx}$ and the Hall conductance $\sigma_{xy}$ of the disordered sample are needed. A simulation of six-terminal Hall bar by using non-equilibrium Green's functions like equation (\ref{3})\cite{Datta} will be numerically expensive. Instead, we adopt the methods which were successful for plotting RG flows\cite{NumScaling}.
The Hall conductance $\sigma_{xy}$ is calculated by a numerical integration of Berry curvature under the Fermi energy $E_F$ using equation (\ref{4}). For a given $E_F$, $\sigma_{xy}$ may not be quantized if some subbands are partly filled. The longitudinal conductance $\sigma_{xx}$ (also in units of $\frac{e^2}{h}$) is calculated
as the Thouless conductance\cite{Thouless,CheckThouless}
\begin{equation}
\sigma_{xx}(E_n)=\pi\rho(E_n)\frac{\partial^2 E_n}{\partial k_x^2}\Big|_{k_x=0} \label{EqThouless}
\end{equation}
with twisted (periodic) boundary condition in $x$ ($y$) direction respectively, where $\rho$ is the density of states, and $E_n(k)$ is the $n$-th subband.
The conductances in Figs. \ref{FigStatistics} and \ref{FigRGFlow} were calculated by these methods.

\section*{Acknowledgements}

Z.-G. S. is grateful to Zhenguo Fu for valuable discussions. This work was
supported by NSFC Nos. 61427901, 11374294 and 11474085, and 973 Program Project No. 2013CB933304.

\section*{Author contributions statement}

Z.-G.S. and Y.-Y.Z. carried out the theoretical calculations and wrote the manuscript with the assistance of J.-T.S. S.-S.L. guided the overall project. All authors reviewed the manuscript.

\section*{Additional information}

\textbf{Competing financial interests:} The authors declare no competing financial interests.


\end{document}